\documentclass[prd,twocolumn,showpacs,preprintnumbers,amsmath,nofootinbib,amssymb,floatfix]{revtex4-1}
\usepackage{graphicx}
\usepackage[sort&compress]{natbib}
\usepackage{subfigure}
\usepackage{amsmath}
\usepackage{amsfonts}
\usepackage{cancel}
\usepackage{lmodern,dsfont}
\usepackage{amssymb}
\begin{document}
\arraycolsep1.5pt
\newcommand{\Ima}{\textrm{Im}}
\newcommand{\Rea}{\textrm{Re}}
\newcommand{\mev}{\textrm{ MeV}}
\newcommand{\be}{\begin{equation}}
\newcommand{\ee}{\end{equation}}
\newcommand{\ba}{\begin{eqnarray}}
\newcommand{\ea}{\end{eqnarray}}
\newcommand{\gev}{\textrm{ GeV}}
\newcommand{\nn}{{\nonumber}}
\newcommand{\dtres}{d^{\hspace{0.1mm} 3}\hspace{-0.5mm}}
\newcommand{\rts}{ \sqrt s}
\newcommand{\non}{\nonumber \\[2mm]}

\title{Scattering of unstable particles in a finite volume: \\
the case of  $\pi\rho$ scattering and the $a_1(1260)$ resonance}

\author{L. Roca$^1$ and E. Oset$^2$}
\affiliation{
$^1$Departamento de F\'{\i}sica. Universidad de Murcia. E-30071, Murcia. Spain\\
$^2$Departamento de F\'{\i}sica Te\'orica and IFIC, Centro Mixto Universidad de Valencia-CSIC,
Institutos de Investigaci\'on de Paterna, Aptdo. 22085, 46071 Valencia,
Spain
}

\date{\today}

\begin{abstract}

  We present a way to evaluate the scattering of unstable particles
quantized in a finite volume with the aim of extracting physical
observables for infinite volume from lattice data. We
illustrate the method with the $\pi\rho$ scattering which generates
dynamically  the axial-vector $a_1(1260)$ resonance. Energy
levels in a  finite box are evaluated both considering the $\rho$ as a stable and
unstable resonance and we find significant differences between both
cases. We discuss how to solve the problem
to get the physical scattering amplitudes in the infinite volume,
and hence phase shifts, from possible lattice results on
energy levels quantized inside a finite box.

\end{abstract}

\maketitle

\section{Introduction}
\label{Intro}

  The determination of hadron spectra is one of the challenging tasks
of Lattice QCD and much effort is being devoted to this problem
\cite{Nakahara:1999vy,Mathur:2006bs,Basak:2007kj,Bulava:2010yg,Morningstar:2010ae,Foley:2010te,Alford:2000mm,Kunihiro:2003yj,Suganuma:2005ds,Hart:2006ps,Wada:2007cp,Prelovsek:2010gm}.
Large pion masses are commonly used in these calculations
\cite{Lin:2008pr,Gattringer:2008vj,Bulava:2010yg,Engel:2010my,Mahbub:2010me,Edwards:2011jj}.
The  ``avoided level crossing'' is usually taken as a signal of a
resonance, but this criteria has been shown insufficient for resonances
with a large width
\cite{Bernard:2007cm,Bernard:2008ax,misha}. A more
accurate method consists on the use of L\"uscher's approach, for resonances with one decay channel, in order to produce phase shifts
for the decay channel from the discrete energy levels in the box
\cite{luscher,Luscher:1990ux}. This method has been recently improved
\cite{misha} by keeping the full relativistic two body propagator
(L\"uscher's approach keeps the imaginary part of this propagator exactly but makes approximations on the real part) and extending the method to two or
more coupled channels. The new method also combines conceptual and
technical simplicity and serves as a guideline for future lattice
calculations. Follow ups of this new practical method have been done in
\cite{mishajuelich} for the application of the J\"ulich approach to
meson baryon interaction and in \cite{alberto} for the interaction of
the $DK$ and $\eta D_s$ system where the $D_{s^*0}(2317)$ resonance is
dynamically generated from the interaction of these particles
\cite{Kolomeitsev:2003ac,Hofmann:2003je,Guo:2006fu,daniel}. The case of
the $\kappa$ resonance in the $K \pi$ channel is also addressed along
the lines of \cite{misha} in \cite{mishakappa}.

  The case of scattering of unstable particles deserves a special care
since in the box one must also discretize the momenta of the decay
products of all the particles. One such system would be the $\pi
\Delta$ system where the $\Delta$ is allowed to decay into $\pi N$. 
The generalization of the work of \cite{misha} to this problem has been
done in \cite{mishaunstable}. 

  The problem of scattering of unstable particles will have to be faced
by the lattice QCD calculations. So far, problems which would require
this treatment have been studied assuming stable particles. This is the
case of the $\pi \rho$ scattering, from where the $a_1(1260)$ resonance
is qualitatively obtained, assuming the $\rho$ to be a stable particle, 
in an actual lattice QCD simulation using the first two levels for a fixed
size of the box \cite{sashatalk}. In the present paper we face directly this problem
and provide the formalism to address it, also for the case of an unstable
$\rho$ resonance. For this purpose recall that in the chiral unitary
approach the axial vector $a_1(1260)$ resonance is dynamically
generated from the interaction of $\pi \rho$ and $\bar{K}K^*$ in 
coupled channels \cite{Lutz:2003fm,luisaxial}, where $\pi \rho$  is
the dominant channel. Then we follow the approach of
Ref.~\cite{luisaxial} and solve the interaction of  $\pi \rho$ within a
box with periodic boundary conditions. These boundary conditions are
imposed on the spectator $\pi$ and on the two $\pi$ that come from the
$\rho$ meson decay. For the $\pi -2\pi$ system we choose the global center of mass (CM)  frame, but the two pions that lead to the $\rho$ in the $\pi
\pi$ loop function are in a moving frame and this forces us to make the
discretization of the levels in this moving frame, a problem which is
well studied in \cite{Rummukainen:1995vs,sachraj}. Furthermore  the
$\rho$ is a p-wave resonance which requires a different method to
obtain the selfenergy than the s-wave resonances. All these problems
will be dealt with in the present paper.

\section{Formalism}

 In the chiral unitary approach the scattering matrix in coupled
channels is given by the Bethe-Salpeter equation in its factorized form

\be
T=[1-VG]^{-1}V= [V^{-1}-G]^{-1},
\label{bse}
\ee
where $V$ is the matrix for the transition potentials between the
channels and 
$G$ 
is a diagonal matrix with the $i^{\rm th}$
element, $G_i$,
given by the loop function of two meson propagators,
a pseudoscalar and a vector meson,
which is defined as 
\be
\label{loop}
G_i=i\,\int\frac{d^4 p}{(2\pi)^4} \,
\frac{1}{(P-p)^2-M_i^2+i\epsilon}\,\frac{1}{p^2-m_i^2+i\epsilon}
\ ,
\ee
where $m_i$ and $M_i$ are the masses of the two mesons 
and $P$ the four-momentum of the global meson-meson system.
Note that in Eq.~(\ref{loop}) we have not considered the
possible widths of the mesons, thus it is only valid for stable mesons.
The unstable mesons case will be addressed in
section~\ref{sec:unstable}.

 For the case of scattering of a pseudoscalar with a vector meson, for
instance $\pi \rho$, $\bar{K}K^*$, as in the present case, the
interaction is taken from the chiral Lagrangians \cite{birse} and has
the form \cite{luisaxial}

\be
V_{PV}=\vec{\epsilon}\cdot\vec{\epsilon}\,' ~V
\ee
where $\vec{\epsilon}$, $\vec{\epsilon}\,'$ are the polarization vectors
of the initial and final vector mesons.  The term
$\vec{\epsilon}\cdot\vec{\epsilon}\,'$ factorizes in all terms of the
Bethe-Salpeter series, $V$, $VGV$, etc., and finally in the T matrix.
Hence, we omit this factor in what follows. 
The 
explicit expression of the potentials, properly projected onto $s$-wave,
is thus \cite{luisaxial}
\ba
V_{ij}(s)=&&\frac{1}{8f^2} C_{ij}
\bigg[3s-(M_i^2+m_i^2+M_j^2+m_j^2)  \nonumber \\
&& -\frac{1}{s}(M_i^2-m_i^2)(M_j^2-m_j^2)\bigg] \ ,
\label{eq:Vtree}
\ea
where  $f=92.5\mev$ is the pion decay constant,
the index $i(j)$
represents the initial (final) $PV$ state in the isospin
basis and
$M_i(M_j)$ and $m_i(m_j)$ correspond  to the masses of 
the initial (final)
vector mesons and initial (final) pseudoscalar mesons, for which we
use an average value for each isospin multiplet. 
 The explicit values
of the numerical  coefficients, $C_{ij}$, can be found in
Ref.~\cite{luisaxial}. For the $\pi\rho$ isovector amplitude, which we
need for the present work,
 $C^{I=1}_{\pi\rho,\pi\rho}=-2$.

The loop function in Eq.~(\ref{loop}) needs to be regularized and
this can be accomplished either with dimensional regularization
or with a three-momentum
cutoff. The equivalence
 of both methods was
shown in Refs.~\cite{ollerulf,ramonetiam}.
In dimensional regularization the integral of Eq.~(\ref{loop}) is evaluated and gives for meson-meson systems
\cite{ollerulf,bennhold}

\begin{eqnarray}
\label{eq:g-function}
&&\nn \mbox G_i(s, m_i, M_i) =  \frac{1}{(4 \pi)^2}
 \Biggr\{
        a_i(\mu) + \log \frac{m_i^2}{\mu^2} \nn\\ +
&&      \frac{M_i^2 - m_i^2 + s}{2s} \log \frac{M_i^2}{m_i^2}\nn \\
     && + \frac{Q_i(\rts)}{\rts}
    \bigg[
         \log \left(  s-(M_i^2-m_i^2) + 2 \rts Q_i(\rts) \right)\nn\\
     && +  \log \left(  s+(M_i^2-m_i^2) + 2 \rts Q_i(\rts) \right)
    \nonumber
  \\
  & &- \log \left( -s+(M_i^2-m_i^2) + 2 \rts Q_i(\rts) \right)\nn\\
&&  - \log \left( -s-(M_i^2-m_i^2) + 2 \rts Q_i(\rts) \right)
    \bigg]
  \Biggr\},
\end{eqnarray}
where $s=E^2$, with $E$ the energy of the system in the center of mass
frame, $Q_i$ the on shell momentum of the particles in the channel $i$,
$\mu$  a regularization scale and $a_i(\mu)$ a subtraction constant
(note that there is only one degree of freedom, not two independent
parameters).

In other works one uses regularization with a cutoff in three momentum
once the $p^0$ integration is analytically performed \cite{npa} and one
gets

\ba
&&G_i=\hspace{-4mm}\int\limits_{|\vec p|<p_{\rm max}}
\frac{d^3\vec p}{(2\pi)^3}\frac{1}{2\omega_1(\vec p)\,\omega_2(\vec p)}
\frac{\omega_1(\vec p)+\omega_2(\vec p)}
{E^2-(\omega_1(\vec p)+\omega_2(\vec p))^2+i\epsilon},
\non 
&&\omega_{1,2}(\vec p)=\sqrt{m_{1,2}^2+\vec p^{\,\,2}}\, ,
\label{prop_cont}
\ea
with $m_1$, $m_2$ corresponding to $m_i$ and $M_i$ of Eq.~(\ref{loop}).

When one wants to obtain the energy levels in the finite box,
 instead of integrating over the
energy states of the continuum with $p$ being a continuous variable
as in Eq.~(\ref{prop_cont}), one must sum over 
the discrete momenta allowed
in a finite box of side $L$ with periodic boundary conditions.
We then have to replace $G$ by 
$\widetilde G={\rm diag}\,(\widetilde G_1,\widetilde G_2)$, where 
\ba
\widetilde G_{j}&=&\frac{1}{L^3}\sum_{\vec p}^{|\vec p|<p_{\rm max}}
\frac{1}{2\omega_1(\vec p)\,\omega_2(\vec p)}\,\,
\frac{\omega_1(\vec p)+\omega_2(\vec p)}
{E^2-(\omega_1(\vec p)+\omega_2(\vec p))^2},
\non 
\vec p&=&\frac{2\pi}{L}\,\vec n,
\quad\vec n\in \mathds{Z}^3 \,
\label{tildeg}
\ea

 This is the procedure followed in \cite{misha}.  The eigenenergies of
the box correspond to energies  that produce poles in the $T$ matrix, 
Eq.~(\ref{bse}), which correspond to zeros of the
determinant of $1-V\widetilde G$.

\section{One channel analysis}

In the present problem the threshold of $\bar{K} K^*$ is above the mass
of the $a_1(1260)$ resonance and, as found in \cite{luisaxial}, the
$\pi \rho$ channel is more important than the $\bar{K} K^*$ one. For
this reason we shall perform the analysis with just the $\pi \rho$
channel, as also done for the lattice calculation of \cite{sashatalk}.

   The one channel problem can be easily solved and is very simple, as
shown in \cite{misha}. The $T$ matrix for infinite volume can be obtained
for the energies which are eigenvalues of the box by 

   \be
T(E)=\left(V^{-1}(E)-G(E)\right)^{-1}= 
\left(\widetilde G(E)-G(E)\right)^{-1} \ . 
\label{extracted_1_channel}
\ee
since $\widetilde G(E)=V^{-1}(E)$ is the condition for the 
$T$ matrix to have a pole for the finite box. 

Hence we find

\begin{align}
T(E)^{-1}=\lim_{p_\textrm{max}\to \infty}
\Bigg[\frac{1}{L^3}\sum_{p_i}^{p_\textrm{max}}I(p_i)
-\int\limits_{p<p_\textrm{max}}\frac{d^3p}{(2\pi)^3} I(p)\Bigg]
\label{tonediff}
\end{align}
where $I(p)$ is the integrand of Eq.~(\ref{prop_cont})

\be
I(p)=\frac{1}{2\omega_1(\vec p)\,\omega_2(\vec p)}
\frac{\omega_1(\vec p)+\omega_2(\vec p)}
{E^2-(\omega_1(\vec p)+\omega_2(\vec p))^2+i\epsilon}.
\label{prop_contado}
\ee

This result is the one obtained in Ref.~\cite{misha} starting with
cutoff regularization and, as proved in Ref.~\cite{misha}, it is
nothing else than L\"uscher formula
\cite{luscher,Luscher:1990ux}, except that Eq.~(\ref{tonediff})
keeps all the terms of the relativistic two body propagator, while in 
L\"uscher's approach one neglects terms in $Re~I(p)$ which are
exponentially suppressed in the physical region, but can become sizable
below threshold, or in other cases when small volumes are used or large
energies are involved.

\section{Generalization to scattering of unstable particles
\label{sec:unstable}
}

\begin{figure}[!h]
\begin{center}
\includegraphics[width=0.9\linewidth]{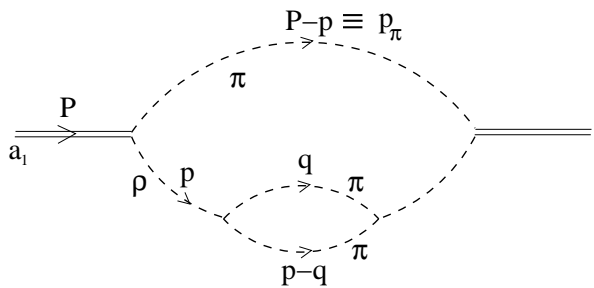}
\caption{The $\pi\rho$ loop diagram considering the $\rho$ meson selfenergy.}
\label{fig:diagram_self}
\end{center}
\end{figure}

In this section we extend the approach to the case where we  have one
unstable particle. We shall work with the case of one channel, but the
generalization to many coupled channels is straightforward. The
consideration of the unstable particle requires to reevaluate the loop
function $G$ of Eq.~(\ref{loop}) by using the dressed meson propagator
including its selfenergy that accounts for the decay channels. This
means substituting

 \be
 \frac{1}{p^2-m^2+i\epsilon}\, \longrightarrow \,
  \frac{1}{p^2-m^2-\Pi(p)},
\label{eq:propself}
 \ee
where $\Pi(p)$ is the meson selfenergy of the unstable particle.
Diagrammatically this means that we must evaluate the loop diagram of
Fig.~\ref{fig:diagram_self}.

In order to calculate this loop function we must first evaluate the
$\rho$ selfenergy, $\Pi(p)$. One knows \cite{Lutz:2003fm,luisaxial} that
differences between the sum and the integral in Eq.~(\ref{tonediff}) in
regions of $p$ far away from the pole of $I(p)$  are exponentially
suppressed in $L$. The sizeable differences stem from regions of $p$
close to the on shell momentum $p_{on}$ where $I(p_{on})$ has a pole.
For this reason we evaluate  $\Pi(p)$ for values of $p$ where the $\pi$
and $\pi \pi$ systems can be placed on shell. This is the same
prescription taken in Ref.~\cite{mishaunstable}. Then we have

\ba \nn
p^2&=& (P- p_\pi)^2=P^2+p_\pi^2-2P p_\pi=s +m_\pi^2-2\sqrt{s}E(\vec p)
\nn\\
&\equiv& M_I^2\equiv s_\rho,
\label{eq:p2srho}
\ea
with $ M_I\equiv \sqrt{s_\rho}$ the invariant mass of the two pion system. We have chosen the CM
for the $\pi\rho$ system and hence $\vec p_\pi=-\vec p$.

Since $\Pi(p)$ is a Lorentz invariant magnitude one can evaluate it in the CM frame of the $\rho$
meson. However, the analogous magnitude $\widetilde \Pi(p)$ in the finite box must take into account
the boundary conditions for the $\pi$ momenta in the moving frame. 

In order to calculate the $\rho\pi\pi$ vertex,
let us consider the standard Lagrangian
for the coupling of one vector to two pseudoscalars  \cite{hidden}

\be
{\cal L}_{\textrm{VPP}}=-i g_V\langle[P,\partial_\mu P]V^\mu \rangle
\ee
where $g_V=M_\rho/2f_\pi$, with $f_\pi=92.5\mev$ the pion decay constant, $P$, $V^\mu$, the
$SU(3)$ matrices of the pseudoscalar and vector mesons and $\langle \dots \rangle$ standing for the $SU(3)$
trace. From this Lagrangian we find
\be
t_{\rho\pi^+ \pi^-}=2\sqrt{2}\,g_V\,\vec q\cdot\vec\epsilon,
\ee
with $\vec q$ taken in the $\rho$ meson rest frame and $\vec \epsilon$ the $\rho$ polarization vector. The
$\rho$ selfenergy is then given by
\ba
-i \Pi(p)&=&\int \frac{d^4q}{(2\pi)^4}\frac{i}{q^2-m_\pi^2+i\epsilon}
\frac{i}{(P-q)^2-m_\pi^2+i\epsilon} \nn \\
&&\times (-i)2\sqrt{2}g_V\,\vec q\cdot\vec\epsilon (-i) 2\sqrt{2}g_V\,\vec q\cdot\vec\epsilon\,'.
\label{eq:Pistart}
\ea
In the $\rho$ rest frame, where we evaluate the $\rho\pi\pi$ vertex,
 we have $q_i \epsilon_i q_j \epsilon'_j$, and
for symmetry reasons we can replace $q_iq_j$ by $\vec q\,^2 \delta_{ij}/3$. Furthermore, in the chiral
unitary approach it is justified to use
 the on-shell approach where the vertices are factorized by their on shell
form. Thus we finally obtain

\be
\Pi(p)=\frac{8}{3} g_V^2 \vec q\,^2_\textrm{on} G_{\pi\pi}(s_\rho) \vec\epsilon\cdot\vec\epsilon\,'
\label{eq:PIpG}
\ee
where $|\vec q_\textrm{on}|=\sqrt{{s_\rho}/{4}-m_\pi^2}$ and $G_{\pi\pi}(s_\rho)$ is the loop
function of Eq.~(\ref{loop})  for two pions. The function $G_{\pi\pi}(s_\rho)$ can be regularized by
means of a cutoff, $q_\textrm{max}$, in the modulus of the
 three-momentum $\vec q$ and its explicit analytic 
expression is \cite{ramonetiam}

\ba
\Pi(s_\rho)&=&\frac{8}{3} g_V^2 \vec q\,^2_\textrm{on} \frac{1}{(4\pi)^2}  \vec\epsilon\cdot\vec\epsilon'
\nn\\
&\times&\left[ \sigma \ln\frac{\sigma r+1}{\sigma r-1}-2\ln\left(\frac{q_\textrm{max}}{m_\pi}(1+r)\right) \right]
\label{eq:PIqmax}
\ea
where $\sigma=\sqrt{1-4m_\pi^2/s_\rho}$ and $r=\sqrt{1+m_\pi^2/q_\textrm{max}^2}$.

In Ref.~\cite{hep-ph/0011096} the $G_{\pi\pi}$ function is evaluated in dimensional regularization.
This is equivalent to removing the divergent part of $\Pi(s_\rho)$ of Eq.~(\ref{eq:PIqmax}) and
substituting it by a subtraction constant which is then constrained by experimental data. 
In Ref.~\cite{hep-ph/0011096}, where the $\rho$ meson is studied within the chiral unitary approach, one
finds that

\be
G_{\pi\pi}^D(s_\rho)=\sigma \ln \frac{\sigma+1}{\sigma-1}+b
\label{eq:gpipisb}
\ee
with
\be
b=-2+d_1^1=-2+\frac{m_K^2}{m_K^2-m_\pi^2}\left( \ln \frac{m_\pi^2}{\mu^2}+\frac{1}{2}\ln
\frac{m_K^2}{\mu^2}+\frac{1}{2}\right)
\ee
and $\mu=m_\rho$.
One way to get the result of Eq.~(\ref{eq:gpipisb}) is to take the limit

\be
G_{\pi\pi}^D(s_\rho)=\lim_{q_\textrm{max}\to\infty}\left[ G_{\pi\pi}(s_\rho)
+\frac{1}{(4\pi)^2}\left( 2\ln \frac{2 q_\textrm{max}}{m_\pi}+b\right) \right].
\label{eq:Gb}
\ee
since $2\ln(2 q_\textrm{max}/m_\pi)$ cancels the divergent
part of the square bracket in Eq.~(\ref{eq:PIqmax}) when
 $q_\textrm{max}\to\infty$.

When we evaluate the selfenergy in the finite box, $\widetilde \Pi(p)$,
we shall use this expression but $G_{\pi\pi}$ will be replaced by the
corresponding discrete sum,  $\widetilde G_{\pi\pi}$. For the numerical
evaluation, the quantity inside the square bracket in the previous
equation 
has to be evaluated for high values of $q_{\rm{max}}$ in order 
to get the convergence.  
However, it oscillates around the convergence value 
for not very large values  of $q_{\rm{max}}$. Hence
an average over the interval
$q_{\rm{max}}\sim[2,2.8]\gev$ numerically gets the convergence value
and thus we take this average in the numerical evaluation, (see the
analogous and further explained reasoning in Ref.~\cite{alberto}).

\section{Discretization of the $\rho$ selfenergy in the moving frame}

The $\rho$ selfenergy in the finite box, $\widetilde \Pi(p)$, is given by Eq.~(\ref{eq:PIpG}),
where now $G_{\pi\pi}$ will be given by   Eq.~(\ref{eq:Gb}) evaluating  $G_{\pi\pi}(s_\rho)$ as a
function of $q_\textrm{max}$ for the finite box. In order to perform this evaluation we must substitute
the $d^3q$ integration implicit in the evaluation of $G_{\pi\pi}(s_\rho)$ in the infinite volume by a
discrete sum over the pion momenta allowed in the finite box with appropriate boundary conditions which
we take to be periodic. The subtlety is that the momenta $\vec q_\textrm{on}$ in Eq.~(\ref{eq:PIpG})
and the variable of integration in Eq.~(\ref{eq:Pistart}) are in the CM of the two pions and the
boundary conditions are in the box, where the pair of pions move with
total momentum $\vec p$. By performing
the $q^0$ integral in Eq.~(\ref{eq:Pistart}) we find in infinite space that

\be
\Pi(p)=8g_V^2 {\vec{q^*}_\textrm{on}}^2\int\limits_{|\vec{q^*}|<q_{\rm max}}
 \frac{d^3q^*}{(2\pi)^3}  
I(q^*) q^*_i \epsilon_i q^*_j \epsilon'_j,
\label{eq:Pistar}
\ee
with $p^2=s_\rho$ and $\vec{q^*}$ the momentum in the CM of the two
pions.

We must write the boost transformation from $q$ to $q^*$. 
By applying
the Lorentz transformation from a
moving frame with momentum $p$
 to a frame where the $\pi\pi$ system is at rest \cite{FTUV-93-43} we find
\be
\vec{q^*}_{1,2}=\vec q_{1,2} + \left[\left(\frac{p^0}{M_I}-1\right)
\frac{\vec q_{1,2}\cdot\vec p}{|\vec p|^2}-\frac{q^0_{1,2}}{M_I}\right]\vec p,
\label{boosteq}
\ee
where $M_I^2=s_\rho=p^2$, ${p^0}^2=M_I^2+\vec p\,^2$ and the subindexes
$1,2,$ represent the two pions of the decay of the $\rho$ meson.
 Demanding that
$\vec{q_1^*}+\vec{q_2^*}=0$ enforces $q_1^0+q_2^0=p^0$
or equivalently $q^{*0}_1+q^{*0}_2=M_I$.
We take for $q^{*0}_1$, $q^{*0}_2$, the on shell pion energy for the
decay of an object of mass $M_I$ at rest into two pions 
\be
q^{*0}_{1,2}=\frac{M_I^2+m_{1,2}^2-m_{2,1}^2}{2M_I}.
\label{q0cm}
\ee
Only this prescription makes $q^{*0}_1=q^{*0}_2$
 for two pions as we should expect for two identical particles in  the
 center of mass.
This provides then the boost for the off shell momenta in
the loop, where $\vec{q}$ is arbitrary but the energy is the on shell
one. Since we need the Jacobian of this transformation, it is useful to
write Eq.~(\ref{boosteq}) in terms of the CM energy of the pion and we
find
\be
\vec{q^*}_{1,2}=\vec q_{1,2} + \left[\left(\frac{M_I}{p^0}-1\right)
\frac{\vec q_{1,2}\cdot\vec p}{|\vec p|^2}
-\frac{q_{1,2}^{*0}}{p^0}\right]\vec p.
\label{boostmisha}
\ee
This equation is the one used in \cite{mishaunstable}.
Furthermore we must substitute $\int \frac{d^3q^*}{(2\pi)^3}$ by
 $\int \frac{d^3q}{(2\pi)^3}\frac{ M_I}{p^0}$, where the factor $\frac{M_I}{p^0}$ is 
the Jacobian of the transformation, with $q$ the $\pi$ momentum in
 the $\pi\pi$ moving frame and then
  $\int \frac{d^3q}{(2\pi)^3}$ becomes
   $\frac{1}{L^3}\sum_{\vec q}$ in the box.
In summary, we must do the substitution
\be
\int \frac{d^3q^*}{(2\pi)^3} \
\longrightarrow \ \frac{1}{L^3}\sum_{\vec q}\frac{ M_I}{p^0},\quad
\vec q=\frac{2\pi}{L}\vec n,\quad \vec n\in \mathds{Z}^3
\ee
for the evaluation of the selfenergy in the box.

When summing over $\vec q$, the integrand
takes the following structure for symmetry reasons:
\be
\sum_{\vec
q}f(\vec{q^*},\vec{q}\,)\epsilon_i{q^*}_i\epsilon'_j{q^*}_j=\epsilon_i\epsilon'_j
(a\delta_{ij}+bp_i p_j)
\label{eq:sumfeps}
\ee
\noindent 
Eq.~(\ref{eq:sumfeps}) is exact for any vector $\vec{p}$ placed along
any of the axis and quite accurate, although not exact for vectors in
other directions. Yet, the term of $b$ is quite small, since the
scale is ${\cal O}((p/m_{\rho})^2)$. Such terms have been systematically
neglected in the approach of \cite{luisaxial} and so do we here. By
contracting Eq.~(\ref{eq:sumfeps}) after removing 
$\epsilon_i\epsilon'_j$
with $\delta_{ij}$ on one hand, and 
with $p_i p_j$ on the other hand, we get two equations from where we
find

\be
a=\frac{1}{2}\sum_{\vec q}f(\vec{q^*},\vec{q}\,)\vec{q^*}^2 (1-\cos^2\theta),
\ee
where $\cos\theta=\vec p\cdot\vec{q^*}/|\vec p||\vec{q^*}|$.
Hence we finally get
\be
\widetilde \Pi(s_\rho)=4g_V^2\frac{1}{L^3}\vec{q^*}^2_\textrm{on}
\sum_{\vec q}^{|\vec{q^*}|<q_{\rm max}}
\frac{ M_I}{p^0\omega_\pi(q^*)} \,
\frac{1-\cos^2\theta}{s_\rho-4 {\omega^2_\pi(q^*)}}
\ee
up to the trivial $\vec\epsilon\cdot\vec\epsilon\,'$ factor.

Finally, in order to get a pole of the $\rho$ propagator in the infinite volume at the physical $\rho$
mass we subtract to $\Pi(s_\rho)$ and $\widetilde \Pi(s_\rho)$ the selfenergy
 $\rm{Re}\, \Pi(m_\rho^2)$.
Thus we replace
\ba
\Pi(s_\rho)\ \longrightarrow \Pi(s_\rho)-\rm{Re}\, \Pi(m_\rho^2),\nn\\
\widetilde \Pi(s_\rho)\ \longrightarrow \widetilde\Pi(s_\rho)-\rm{Re}\, \Pi(m_\rho^2).
\ea

\section{Inclusion of the $\rho$ selfenergy in the finite box and infinite volume}

We come back to the original problem of the $\pi\rho$ interaction with the $\rho$ dressed by its
selfenergy. In the approximation that we did to write $\Pi_\rho$ for the case where the three
intermediate pions are placed on shell, the variable $s_\rho$ depends on $\vec p$, see
Eq.~(\ref{eq:p2srho}). Hence, the $p^0$ integration of Eq.~(\ref{loop}) when we replace the vector meson
propagator as in Eq.~(\ref{eq:propself}) can be performed in the same way as with two free propagators
and we obtain
\ba
G_{\pi\rho}(E)&=&\int\limits_{|\vec p|<p_{\rm max}} \frac{d^3 p}{(2\pi)^3}
\frac{1}{2\omega_1\omega_2}  \nn\\
&\times&\frac{\omega_1+\omega_2}
{E^2-(\omega_1+\omega_2)^2
-\frac{\omega_1+\omega_2}{\omega_2}\Pi(s_\rho)}
\label{eq:Gfinal}
\ea
where $\omega_1\equiv\omega_1(\vec p)$ and
 $\omega_2\equiv\omega_2(\vec p)$ are the $\pi$ and $\rho$ on shell
energies, and we have kept the leading term in $\Pi/\omega_2^2$, which is a small quantity
close to the $\rho$ on-shell from where the finite volume corrections stem mostly.

For the box we substitute $G_{\pi\rho}$ by $\widetilde G_{\pi\rho}$ given by
\ba
\widetilde G_{\pi\rho}(E)&=&\frac{1}{L^3}\sum_{\vec p}^{|\vec p|<p_{\rm max}}
\frac{1}{2\omega_1\omega_2} \nn\\
&\times&\frac{\omega_1+\omega_2}
{E^2-(\omega_1+\omega_2)^2
-\frac{\omega_1+\omega_2}{\omega_2}\widetilde\Pi(s_\rho)}.
\label{eq:Gboxfinal}
\ea

In one channel the scattering matrix in infinite volume is given by

\be
T=\frac{1}{V^{-1}-G_{\pi\rho}}
\label{eq:TVGinf}
\ee
and for the finite box
\be
T=\frac{1}{V^{-1}-\widetilde G_{\pi\rho}}.
\ee

The eigenenergies of the unstable $\pi\rho$ system in the box are given by the energies that satisfy
\be
V^{-1}=\widetilde G_{\pi\rho}.
\label{eq:VeqG}
\ee

\begin{figure}[!h]
\begin{center}
\includegraphics[width=0.9\linewidth]{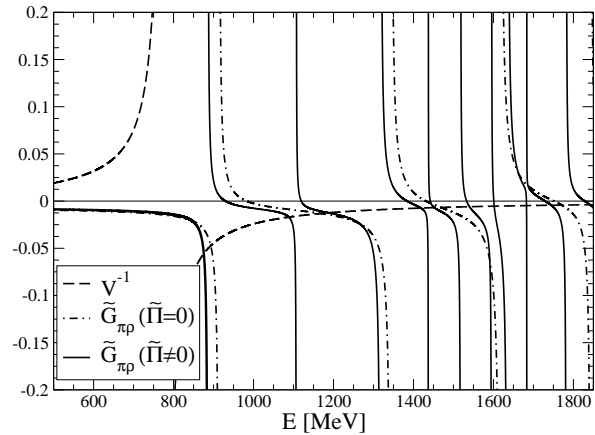}
\caption{Loop function in the box $\widetilde G_{\pi\rho}$ (solid line) and $V^{-1}$ (dashed line)
for $L=2 m_\pi^{-1}$ and
$p_\textrm{max}=1\gev$. The dashed dotted line corresponds to
 the case with stable $\rho$, $\widetilde\Pi(s_\rho)$=0.}
\label{fig:GdeE}
\end{center}
\end{figure}

In Fig.~\ref{fig:GdeE} we plot $V^{-1}$ and $\widetilde G_{\pi\rho}$ as
a function of $E$ for $L=2 m_\pi^{-1}$ and $p_\textrm{max}=1\gev$. The
$\widetilde G_{\pi\rho}$ plot  is very different to a typical
meson-meson loop function in infinite volume. It shows clear poles
coming from zeros in the denominator of Eq.~(\ref{eq:Gboxfinal}) and
from poles of $\widetilde\Pi(s_\rho)$. These poles of 
$\widetilde G_{\pi\rho}$ are not present in the
infinite volume since in the integration the poles of
the integrand provide an imaginary part
to the loop function but not a pole after performing the integration.
 However, this is not the case in the finite box since
we do not have an integral but a summation. The intersection between 
both plots of Fig.~\ref{fig:GdeE} just
 provides the $\pi\rho$ scattering
eigenenergies in the box. 

It is interesting to note that the spectra obtained is also qualitatively different to the one obtained for the stable $\rho$ which would be given by the intersection of $V^{-1}$ and $\widetilde G_{\pi\rho}$ for the case of the stable $\rho$ (dashed-dotted line in the figure). We can see that for stable $\rho$ one has $\widetilde G_{\pi\rho}$ going to infinity when the energy $E$ approaches one of the free energies of the $\pi \rho$ system in the box. Then $V^{-1}$ cuts $\widetilde G_{\pi\rho}$ only once in between two neighboring asymptotes. When we discretize the $\pi \pi$ system, $\widetilde\Pi(s_\rho)$ becomes infinite for the discrete energies of the moving $\pi \pi$ system in the box. With $\widetilde\Pi(s_\rho)$ becoming infinite, with plus and minus sign, in the denominator of Eq.~(\ref{eq:Gboxfinal}), independently of the value of $E^2-(\omega_1+\omega_2)^2$, close to the pole of $\widetilde\Pi(s_\rho)$ there will be an energy where the denominator will vanish, leading to a pole of  $\widetilde G_{\pi\rho}$. Thus, we get asymptotes of $\widetilde G_{\pi\rho}$ for values of E close to the free eigenenergies of $\pi \rho$ and also close to the free eigenenergies of the moving $\pi \pi$. We observe that in between two asymptotes corresponding to the $\pi \rho$ free eigenenergies (dashed dotted lines) one new asymptote has appeared corresponding to a free $\pi \pi $ eigenenergy of the $\pi \pi$ moving frame. As a result of it, the line $V^{-1}$ cuts now two lines corresponding to the unstable $\rho$ and only one corresponding to the stable $\rho$.  If we go to higher energies we observe that in between the next two asymptotes corresponding to the free eigenenergies of $\pi \rho$ there are now three extra asymptotes corresponding to free eigenenergies of $\pi \pi$ in the moving frame. This leads now to four eigenenergies of the interacting $\pi \rho$ system in the box when we cut these lines with $V^{-1}$. There is, thus, a proliferation of eigenenergies as a consequence of considering the $\rho$ as an unstable state.

\section{Results}

\subsection{Energy levels in the box}

\begin{figure}[!h]
\begin{center}
\includegraphics[width=0.9\linewidth]{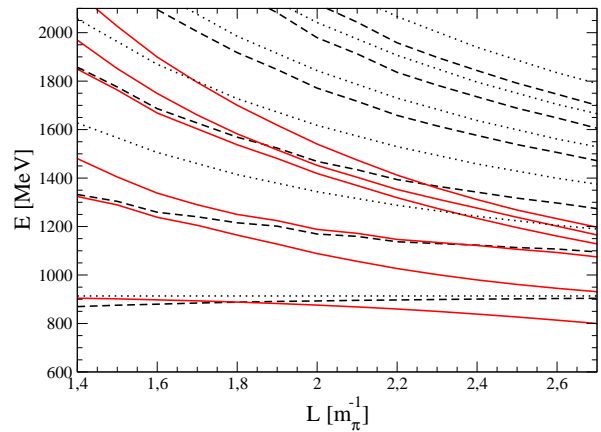}
\caption{The first six energy levels as a function of the cubic box size $L$ for stable $\rho$-meson
($\widetilde \Pi =0$) (dashed lines) and for unstable $\rho$-meson (solid lines) using 
$p_\textrm{max}=1\gev$. The dotted lines indicate the free $\pi \rho$ energies of the box for comparison.}
\label{fig:EdeL}
\end{center}
\end{figure}

We first show in Fig.~\ref{fig:EdeL} the solutions of
Eq.~(\ref{eq:VeqG}) which represent the  energy levels for different
values of the cubic size, $L$. The solid lines represent the full
calculation, considering the unstable $\rho$-meson via the
consideration of its selfenergy in the box, $\widetilde \Pi$ in
Eq.~(\ref{eq:Gboxfinal}), and the dashed lines are the same calculation
but setting $\widetilde \Pi=0$ in Eq.~(\ref{eq:Gboxfinal}), 
{\emph i.e.},
considering the case of stable $\rho$ meson. There is an
infinite number of levels and we have plotted the first six ones
for illustration. This distribution of energy levels depends on
the cutoff $p_\textrm{max}$ and in particular this figure has been
evaluated using $p_\textrm{max}=1\gev$ which is of the order of the
cutoff used in Ref.~\cite{luisaxial} to get dynamically the
axial-vector resonances.

It is worth noting the significant difference between the consideration of the unstable $\rho$ meson in
comparison to the stable case. The main difference is that the levels tend to decrease and shrink as
$L$ increases and that there are extra levels between those already present in the stable case, for
instance between the second and third levels. This
latter feature comes from the appearance of extra poles in $\widetilde G_{\pi\rho}$ due to the poles in 
$\widetilde \Pi$ and this produces extra intersections with 
the smooth function $V^{-1}$
as explained above. This different behavior between the
stable and unstable analysis must be considered as a caveat for
lattice calculations that only consider the stable case.

\subsection{Inverse problem: getting $\pi\rho$ amplitudes and phase shifts from 
{\it lattice-like} data}

By ``inverse problem'' we refer to the problem of getting the actual
scattering amplitudes (and hence by-product magnitudes like
phase shifts) in the infinite space from data consisting of points over
the energy levels in the box in the $E$~vs.~$L$ plots,  which is what a
lattice calculation would provide. In our case we can ``synthetically''
simulate this {\it lattice-like} data from our model generating points in the
levels of Fig.~\ref{fig:EdeL}. We will call this data {\it lattice-like}
set although in the present work it is generated from our model for
illustrative  purposes.

\begin{figure}[!h]
     \centering
      \subfigure[]{\label{fig:EdeLdata_a}
          \includegraphics[width=.9\linewidth]{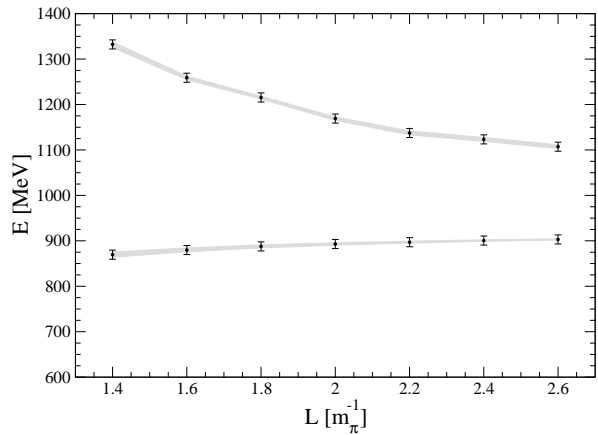}} \\
      \subfigure[]{\label{fig:EdeLdata_b}
          \includegraphics[width=.9\linewidth]{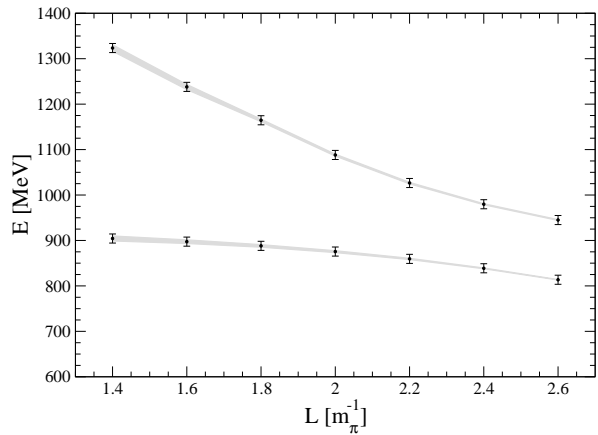}}
   \caption{Synthesized {\it lattice-like} data considered
 in order to illustrate the inverse problem methods.
 Stable $\rho$ case: (a).
 Unstable $\rho$ case: (b).
The error band represent the result of the fit in the {\it fit method}
described in the text.}
\label{fig:EdeLdata}
\end{figure}

 In Fig.~\ref{fig:EdeLdata} we represent by error bars  the set
generated for a particular election of $L$ values for the stable, 
\ref{fig:EdeLdata_a},
and unstable, \ref{fig:EdeLdata_b}, $\rho$ cases, to which
 we have assigned
a reasonable  error of
$10\mev$.  The meaning of the shadowed error bands will be explained later
in this section when explaining the {\it fit method} \cite{misha},
hence they must be forgotten for the moment.
 We are
aware that it is difficult for a present lattice calculation to get
such quantity of points as considered in Fig.~\ref{fig:EdeLdata},
but the method is equally valid for less points and we have chosen such
a set just to illustrate more clearly the method. In an actual inverse
problem the {\it lattice-like} generated set would just be replaced by
actual lattice results.

In Refs.~\cite{misha,mishajuelich,alberto,mishakappa}
several methods were suggested to solve the inverse problem, mostly based on
fitting the potential $V$ to reproduce {\it lattice-like} data
analogous to those in Fig.~\ref{fig:EdeLdata}.
In the following we will call this method
{\it fit method}.
We will also discuss later the {\it fit method}
for the present problem but first we want to propose a different
strategy which does not require to assume a specific
shape of the potential $V$, unlike the {\it fit method}.
In the following we will call this new method {\it direct method} 
\cite{misha}.
The idea of the {\it direct method}
is to evaluate directly the $\pi\rho$ amplitude using the expression
  \be
T_{\pi\rho}(E)=\frac{1}{\widetilde G_{\pi\rho}(E)-G_{\pi\rho}(E)}. 
\label{extracted_1_channel2}
\ee
in the $p_\textrm{max}\to\infty$ limit
with $G_{\pi\rho}(E)$ and $\widetilde G_{\pi\rho}(E)$ from Eqs.~(\ref{eq:Gfinal}) and
 (\ref{eq:Gboxfinal}) for the energies of the points in Fig.~\ref{fig:EdeLdata}.
Recall that Eq.~(\ref{extracted_1_channel2}) is valid only for the energies solution 
of Eq.~(\ref{eq:VeqG}).
  Note that despite the fact that $G_{\pi\rho}(E)$ 
 and $\widetilde G_{\pi\rho}(E)$ are divergent in the limit $p_\textrm{max}\to\infty$, the difference
$ \widetilde G_{\pi\rho}(E)-G_{\pi\rho}(E)$ which appears in Eq.~(\ref{extracted_1_channel2}) is
convergent. Therefore Eq.~(\ref{extracted_1_channel2}) is cutoff independent. 
This is definitely a
non-trivial result an illustrates one of the strong points of the method to
solve the inverse problem. For practical numerical evaluations we have checked that considering an
average  in the interval
$p_{\rm{max}}\sim[1.5,2.5]\gev$ we get the same numerical result than considering 
$p_\textrm{max}\to\infty$. As shown in \cite{misha}, Eq. (\ref{extracted_1_channel2}) is a different and practical way of writing L\"uscher's equation, taking into account the full relativistic two body propagators.

\begin{figure}[!h]
     \centering
      \subfigure[]{
          \includegraphics[width=.9\linewidth]{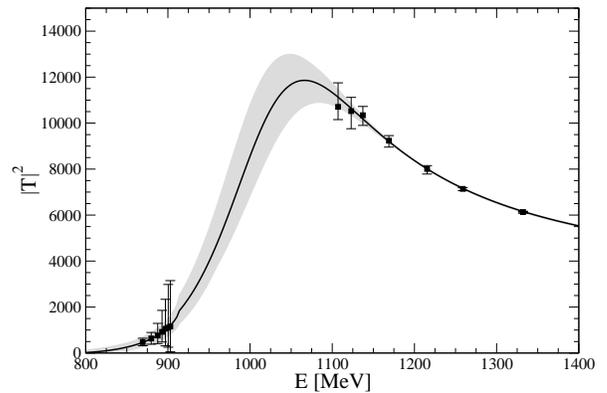}} \\
 
      \subfigure[]{
          \includegraphics[width=.9\linewidth]{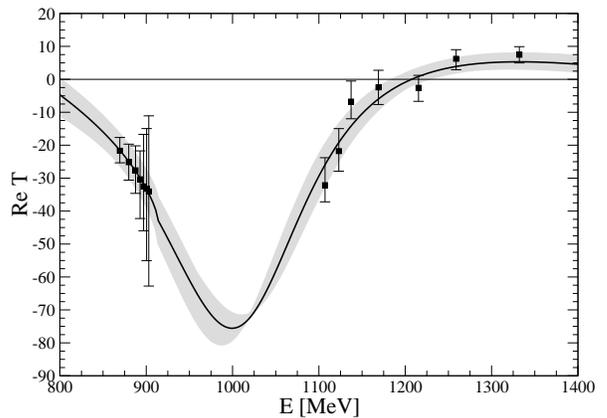}} \\
       \subfigure[]{
          \includegraphics[width=.9\linewidth]{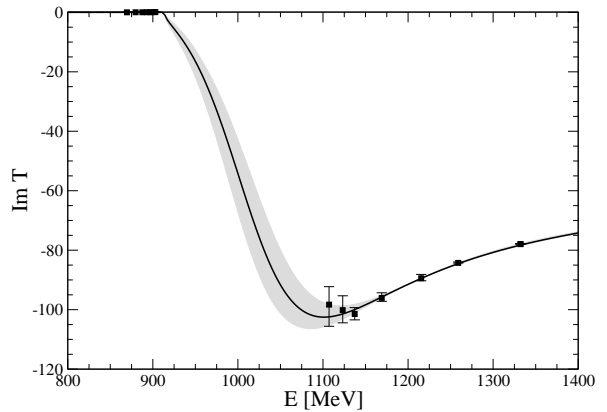}} \\
   \caption{The $\pi\rho$ scattering amplitude solution of the inverse
   problem for the stable $\rho$ case. Horizontal errors of $\pm 10~ MeV$ in the energy axis have to be also assumed. 
 Square points with error bars: {\it direct method}. 
 Solid line with error band: {\it fit method}  }
     \label{fig:TsdirectG0}
\end{figure}

\begin{figure}[!h]
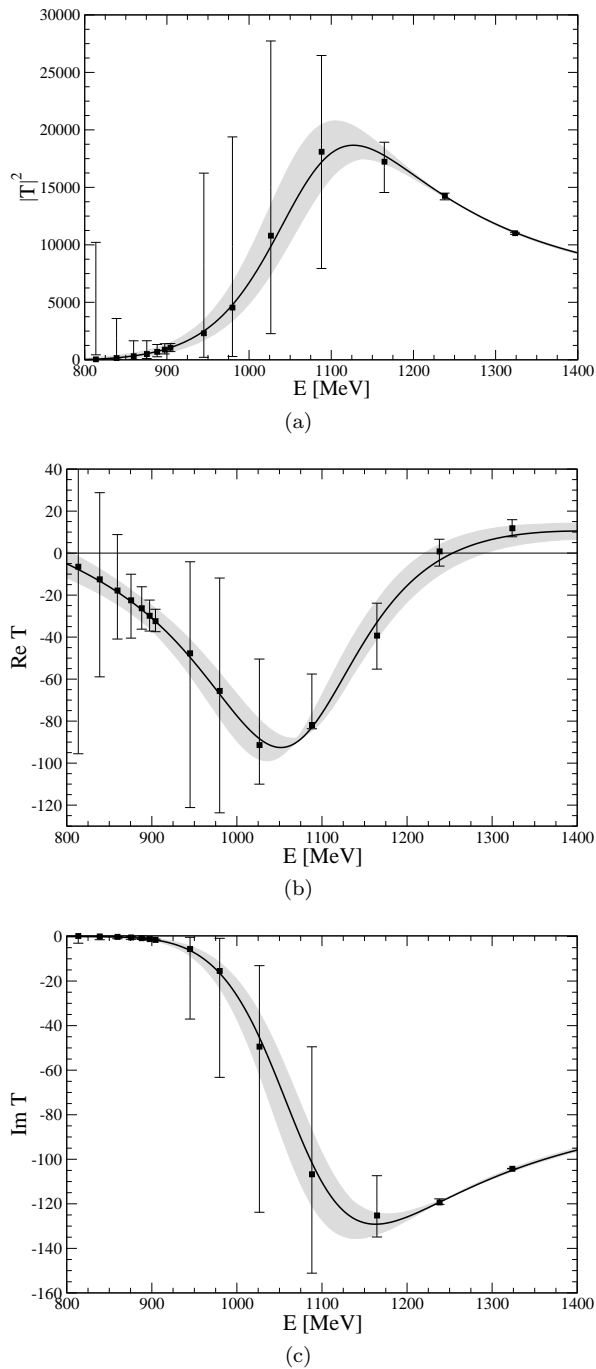

     \centering
      \subfigure[]{
          \includegraphics[width=.9\linewidth]{figure6a.eps}} \\
      \subfigure[]{
          \includegraphics[width=.9\linewidth]{figure6b.eps}} \\
       \subfigure[]{
          \includegraphics[width=.9\linewidth]{figure6c.eps}} \\
   \caption{Same as Fig.~\ref{fig:TsdirectG0} but for unstable $\rho$
   meson }
     \label{fig:TsdirectG}
\end{figure}

In Figs.~\ref{fig:TsdirectG0} and \ref{fig:TsdirectG} we represent by
square points with error bars the
$\pi\rho$ amplitude  (modulus squared, real part and imaginary part)
for the stable (Fig.~\ref{fig:TsdirectG0}) and unstable
(Fig.~\ref{fig:TsdirectG}) $\rho$ cases for the 
{\it direct method}. (The solid line and shadowed error bands represent the
solution of the {\it fit method} which will be explained later on).
 The central points of the solution of the {\it direct method} are
obtained with the central values of the {\it lattice-like} set and the error
bars are evaluated by varying the energies within the $10\mev$ errors given
in Fig.~\ref{fig:EdeLdata}.  For the stable case there are no data
points between 900 and 1100$\mev$ since, as can be seen in
Fig.~\ref{fig:EdeL},  one should go to very high values of the
size of the box in order to get a good resolution in this energy
region.  It is worth noting that this method produces large errors for
certain energies, particularly in the case of the unstable $\rho$. This is due to the fact that  $\widetilde
G_{\pi\rho}(E)$ is usually very steep close to the energy values of the
levels (see  Fig.~\ref{fig:GdeE}) and, hence, small variations in $E$
provides large variations in $\widetilde G_{\pi\rho}(E)$. This feature
can be seen in the $\widetilde G_{\pi\rho}$ vs. $E$ plot in
Fig.~\ref{fig:GdeE} where  it is visible that the crossing between the
$V^{-1}$ and  $\widetilde G_{\pi\rho}$ lines usually occur close to
poles  of $\widetilde G_{\pi\rho}$ and hence in an energy region where
$\widetilde G_{\pi\rho}$ changes rapidly. Anyway, a clear resonance
shape corresponding to the $a_1(1260)$ is visible as a peak in $|T|^2$. Note, however, that
the shape is far from being a Breit-Wigner. 
The chiral unitary approach
in which our {\it lattice-like} data set is generated, provides not only
poles but the full scattering amplitude in the complex plane, in
particular also in the real axis, and generates the
possible background
besides the pole, {\emph i.e.}, provides the actual shape of the amplitude. 
 Actually, for the $a_1(1260)$ there is a very strong background which
 distorts the shape from a Breit-Wigner.
The reason is that the amplitude is zero
 at around $E\simeq M_\rho$
since the potential
$V$ is zero around that energy, 
as can be easily seen just looking for zeroes of Eq.~(\ref{eq:Vtree})
for the $\pi\rho$ case.
However, the pole contribution itself has a large strength at this
energy.
 This means that there must be a very
strong background in order to cancel at that energy the tail of the
Breit-Wigner shape coming from the pole in order to produce the zero.
(See a more extended discussion on this zero 
in Ref.~\cite{arXiv:0911.1235}).

\begin{figure}[!h]
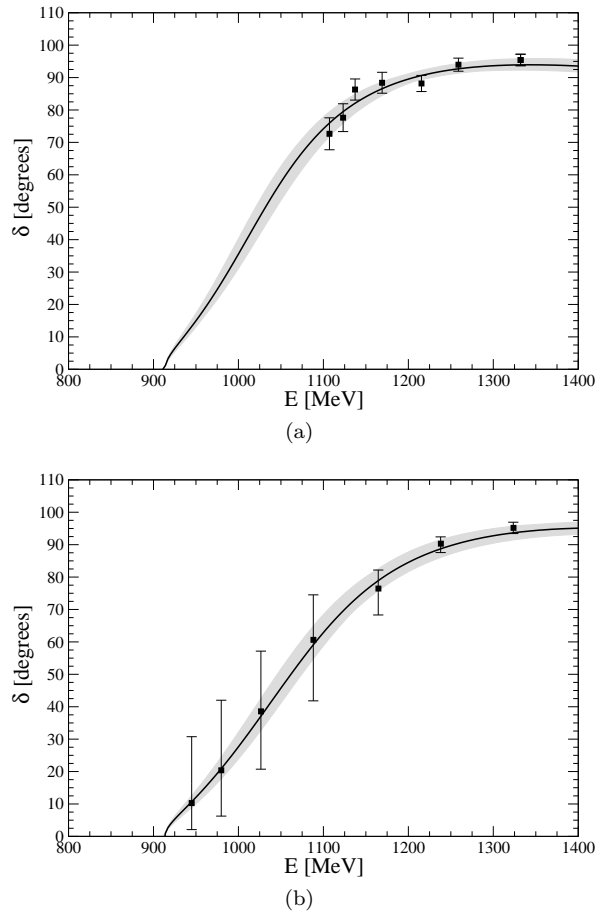

     \centering
      \subfigure[]{ \label{fig:phaseshiftsa}
          \includegraphics[width=.9\linewidth]{figure7a.eps}} \\
      \subfigure[]{ \label{fig:phaseshiftsb}
          \includegraphics[width=.9\linewidth]{figure7b.eps}}
   \caption{Phase shifts obtained from the solution of the inverse
   problem for the stable (a) and unstable (b) cases.Horizontal errors of $\pm 10~ MeV$ in the energy axis have to be also assumed. 
 Square points with error bars: {\it direct method}. 
 Solid line with error band: {\it fit method}}
     \label{fig:phaseshifts}
\end{figure}

From the scattering amplitudes we can get the phase shift, $\delta$,
that is well
defined for stable particle an which in our normalization it is related
to the amplitude via

\ba
&&S=e^{2i\delta}=1-i\frac{p}{4\pi E}T \nn \\
&\Longrightarrow& T(E)=-\frac{8\pi E}{p}\frac{1}{\cot \delta-i}.
\label{eq:defas}
\ea
where $p=\sqrt{(E^2-(m_\rho+m_\pi)^2)(E^2-(m_\rho-m_\pi)^2)}/(2E)$ 
is the CM momentum.
Note that 
Eq.~(\ref{eq:defas}) is only well defined for stable scattering 
particles. However, for the sake of comparison between the stable and
unstable case we also use, by definition for the unstable case,
the same equation using for the $\rho$ meson
mass the physical value, but being aware that in the unstable case this
is not a well defined observable and it is just a theoretical
mathematical exercise.
The solutions of the inverse problem for the phase shifts using
 the {\it direct method} 
are shown by the squares and error bars in 
Fig.~\ref{fig:phaseshifts} for the stable, \ref{fig:phaseshiftsa},
 and unstable, \ref{fig:phaseshiftsb}, cases. (Again the 
solid line and shadowed error bands represent the
solution of the {\it fit method} explained below.)

 It is interesting to note that the phase shifts that we have obtained are quite different from those of a Breit Wigner, where the phase shift would go from zero to 180 degrees passing through 90 degrees at the pole of the resonance. We see that the phase shifts stabilize around 90-100 degrees at high energies around 1400 MeV. On the other hand, the shape of $|T^2|$ in Fig. \ref{fig:TsdirectG0} is typical of a resonance, and bumps like that would be used to identify the resonance experimentally. To the light of this, and the comments above that the $\pi \rho$ amplitude has a strong background at low energies, one might question the procedure used in 
\cite{sashatalk} where the shape of the amplitude is constructed from a calculated phase shift below threshold and another one above, assuming that one has a Breit Wigner amplitude. On the other hand, given the peculiar shape of the amplitude predicted here, the evaluation of phase shifts at physical energies in the energy range of our Fig. \ref{fig:phaseshifts} would be most welcome. In this sense, it is interesting to note that the approach of \cite{sashatalk} produces $\delta$ around 90 degrees in the region around 1430 MeV, where we find the phase shift stabilizing around that value (see Fig.~\ref{fig:phaseshifts}).

Let us now apply the other method
already introduced in 
 Refs.~\cite{misha,alberto}
  to solve the inverse problem
based on fitting the potential, {\it fit method}.

The shape of the lowest order $\pi\rho$ potential, $V$,
based on Eq.~(\ref{eq:Vtree}) which comes essentially 
from chiral symmetry, can
be written in the form
\cite{luisaxial}:

\be
V=a'+b's+\frac{c'}{s}.
\label{eq:Vabcp}
\ee
In order to make the coefficients $a$, $b$ and $c$ adimensional
 and of
natural order 1 we redefine the previous equation as
\be
V\equiv\frac{-1}{4 f_\pi^2}\left[m_R^2 a+b (s-m_R^2)- \frac{m_\rho^2
}{s}c\right]
\label{potchiral}
\ee
with $m_R=1.2\gev$. Actually in Ref.~\cite{luisaxial}
 $a$, $b$ and $c$ were
 around 2, 1   and 3 respectively for the present channel, as can be
 obtained from Eq.~(\ref{eq:Vtree}). One should note that the expression
of Eq.~(\ref{potchiral}) is the one that one has in the chiral unitary
approach. In this sense, the fit that we perform provides as a best solution
the results of the chiral unitary approach with $\chi^2_{\rm{min}}=0$.
One can give one self some freedom to have other possible potentials.
When this is done what one finds is that the best solution is
essentially the same but the uncertainties are somewhat larger
\cite{mishakappa}.

 Next, we assume that lattice data are provided by our synthetic
 {\it lattice-like} set, points with error bars  in
  Fig.~\ref{fig:EdeLdata}.
   Then we fit these points with
 the solutions coming from Eq.~(\ref{eq:VeqG}) in order to get the
best $a$, $b$ and $c$ parameters, which produce the minimum $\chi^2$,
$\chi^2_{\rm{min}}$. 

The $\widetilde G_{\pi\rho}(E)$ function is of course dependent on the
cutoff $p_\textrm{max}$ and, therefore, also are the parameters fitted.
However, the amplitude $T(E)$ obtained from Eq.~(\ref{eq:TVGinf}) should
be
independent of this cutoff and therefore the inverse method does not
require the knowledge or assumption of any particular cutoff, 
{\emph i.e.},
changing the cutoff value would produce different values for  $a$,
$b$ and $c$ but the amplitudes and observables derived from them would
be the same. We have checked numerically that this is indeed the case
within errors. This feature is not novel, it was already observed in \cite{misha}
  and is related to the behavior of amplitudes within the renormalization
group method, as used for instance in Quantum Mechanics in \cite{bengt}.

 In Fig.~\ref{fig:EdeLdata}
  we show with the solid line
  the result of the fit to the energy levels using
 for the fit the cutoff value $p_\textrm{max}= 1\gev$.
 The shadowed band represents the assigned errors obtained by varying
 the fitted parameters such that $\chi^2\le \chi^2_{\rm{min}}+1$.
   From Fig.~\ref{fig:TsdirectG0} till Fig.~\ref{fig:phaseshifts} we
 represent by the solid line and error bands the results of the 
 {\it fit method} for the $\pi\rho$ amplitude and phase shift.
Note that the {\it fit method} gives generally smaller errors than 
the {\it direct method} since it considers a global fit to all the points
of the 
lattice-like set and does not suffer from the pathological error
sensitivity discussed above for the 
{\it direct method}.
 However, it has the drawback of having to assume a
shape of the 
potential, $V$,  like in Eq.~(\ref{eq:Vabcp}).

It is remarkable that in spite of the proliferation of eigenenergies in the
case of the unstable $\rho$ and the different shapes of the two levels in Fig.
\ref{fig:EdeLdata} for the stable and unstable $\rho$, the amplitudes and the
phase shifts in Figs. \ref{fig:TsdirectG0}, \ref{fig:TsdirectG} and
\ref{fig:phaseshifts} are rather similar. We observe then that the effects of
considering the $\rho$ unstable are rather moderate
in the previous analysis. This is not the case if we start from data 
generated
with unstable $\rho$ but the inverse problem is analyzed with stable $\rho$,
as will be explained at the end of this section.
It is interesting to
observe that the proliferation of levels in the case of the
unstable $\rho$ has made the region of energies between threshold and 1400
MeV accessible, while for the stable $\rho$ the first accessible physical
energies are around 1100~MeV. The price one pays for having these energies
available in the unstable $\rho$ case is a bigger uncertainty in the
reconstruction of the phase shifts as seen in Fig. \ref{fig:phaseshifts}.


\begin{figure*}[h]
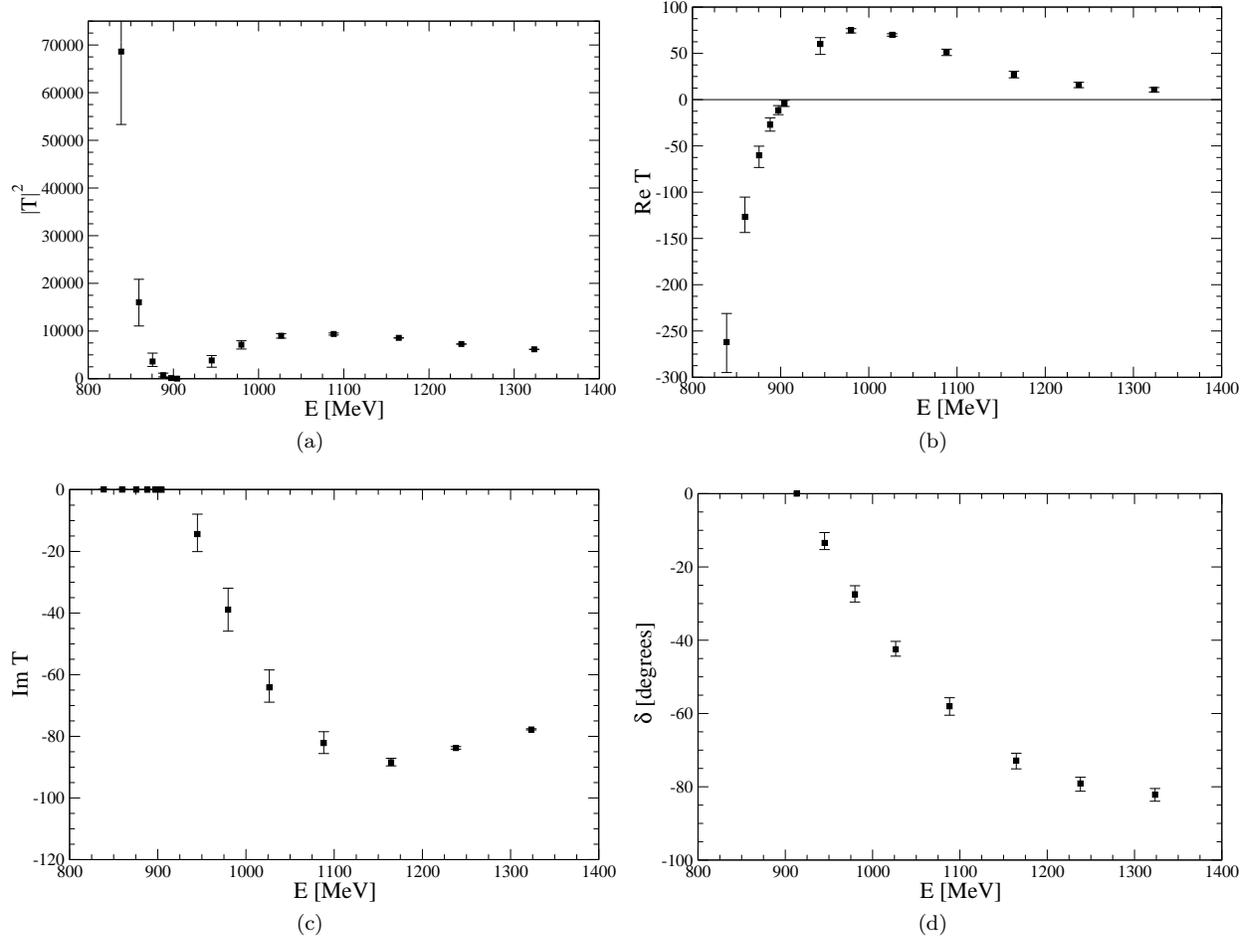

     \centering
      \subfigure[]{
          \includegraphics[width=.45\linewidth]{figure8a.eps}} 
       \subfigure[]{
          \includegraphics[width=.45\linewidth]{figure8b.eps}} \\
       \subfigure[]{
          \includegraphics[width=.45\linewidth]{figure8c.eps}} 
       \subfigure[]{
          \includegraphics[width=.45\linewidth]{figure8d.eps}} \\
   \caption{The $\pi\rho$ scattering amplitude and phase shift 
   solution of the inverse
   problem with the {\it direct method}  with stable $\rho$ case
 but starting from data generated with  unstable $\rho$.
   }
     \label{fig:Tdefamixed}
\end{figure*}

\begin{figure}[!h]
\begin{center}
\includegraphics[width=0.9\linewidth]{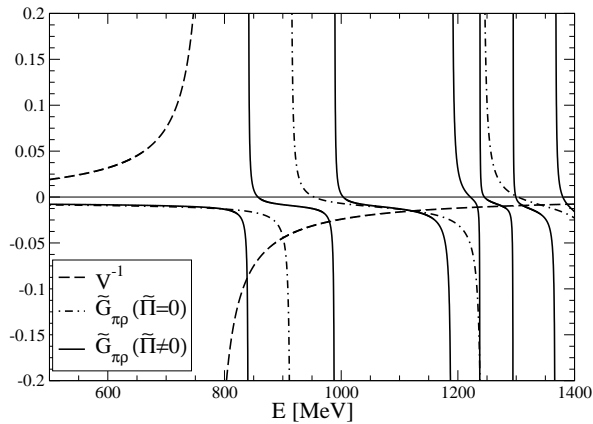}
\caption{Same as Fig.~\ref{fig:GdeE} but for $L=2.4\, m_\pi^{-1}$. }
\label{fig:GdeE24}
\end{center}
\end{figure}

At this point let us discuss a different analysis which may be closer to what
a lattice calculation implements (see, however, some technical observations at the end of this section). In a real
lattice calculation the generated data should be closer to those
in  Fig.~\ref{fig:EdeLdata_b}, unstable $\rho$ case, since starting from
quarks in the lattice and imposing on them the boundary conditions would naturally
lead to boundary conditions in all the pions, the spectator and those coming from the $\rho$ decay.  However, for simplicity, one may be tempted to do the 
subsequent analysis using the model for the stable $\rho$. In view of this, we are
going to see what happens if one starts from a lattice-like set generated with
the unstable model but the inverse problem 
is analyzed considering the stable case.
Let us first focus on the direct method. Thus we  evaluate now
Eq.~(\ref{extracted_1_channel2}) for the case of stable $\rho$-meson but for
the energies shown in Fig.~\ref{fig:EdeLdata_b}. The results for the
amplitude and phase shift are shown in 
Fig.~\ref{fig:Tdefamixed}. The result is very far from reality. The real part of the amplitude blows-up for low
energies and it has little to do with
the results of Figs.~\ref{fig:TsdirectG0} or
\ref{fig:TsdirectG}. The phase shift is also different to 
Fig.~\ref{fig:phaseshifts}, even the sign. Conceptually there are no
deep reasons why  the results should be so different, but numerically the  method
is pathological for the following reasons. Let us focus on one energy where
the extracted amplitude is very large, for instance the point of the lower
energy band at $L=2.4\,m_\pi^{-1}$, for which $E=839\mev$, (see 
Fig.~\ref{fig:EdeLdata_b}). Let us look now at
Fig.~\ref{fig:GdeE24}, where we show the same plots as in Fig.~\ref{fig:GdeE} but for 
$L=2.4\,m_\pi^{-1}$, (Fig.~\ref{fig:GdeE} was evaluated  for
$L=2\,m_\pi^{-1}$). The energy $E=839\mev$ is indeed the first crossing point
between $V^{-1}$ (dashed line) and $\widetilde G_{\pi\rho}(\widetilde\Pi\ne
0)$ (solid line). However, at this energy the value of  $\widetilde
G_{\pi\rho}(\widetilde\Pi= 0)$ (dashed dotted line) is very different, about
a factor six smaller.  In  Eq.~(\ref{extracted_1_channel2}), $G_{\pi\rho}$
is similar both for stable and unstable case and has a small value at this
energy, (about -0.02). Therefore the amplitude evaluated with $\widetilde
G_{\pi\rho}(\widetilde\Pi= 0)$ is much larger than the one evaluated with
$\widetilde G_{\pi\rho}(\widetilde\Pi\ne 0)$. On the other hand, the extra
poles in $\widetilde G_{\pi\rho}(\widetilde\Pi\ne 0)$ coming from the 
eigenenergies of the $\pi\pi$ system as explained above, make the plot of 
$\widetilde G_{\pi\rho}(\widetilde\Pi\ne 0)$ very different from $\widetilde
G_{\pi\rho}(\widetilde\Pi= 0)$. Thus, close to an extra pole
this numerical analysis is also inappropriate, since one would 
be misidentifying the poles responsible for the eigenvalues using the two procedures. 
 These pathological behaviors of the 
$G_{\pi\rho}$ functions within the box  are essentially due to the abruptness
produced by the poles of the $G$ function in the box and it makes very difficult to
extract reliable information from this analysis. 
We have also implemented the {\it fit method} in the last analysis,
with stable
$\rho$  but starting from data created with unstable
$\rho$. In this case, no good fit is obtained since with this shape of the
potential it is not possible to generate a curve decreasing with $L$ as the
lower level in Fig.~\ref{fig:EdeLdata_b} does. Actually the best fit
 has
a $\chi^2\sim 200$ and produces the pole for the $a_1$ 
in the real axis below
threshold which makes also this analysis unreliable.
The main conclusion of the analysis using the stable case but starting from
data generated using unstable $\rho$ meson is that 
the inverse method
should be as close as possible to reality, which in this context means to
consider the unstable $\rho$ meson in the analysis of the data. 
Otherwise unreliable results could be
obtained due to the abrupt shape of the $G$ functions in the box.
  
In practice, the above mentioned problem could not show up in some actual lattice calculations, like in \cite{sashatalk}, if large pion masses are used. Indeed, we have seen that with the pion mass $m_{\pi}=266 MeV$, used in \cite{sashatalk}, the ground state level is much less affected than in the case shown here, and the extra level originating from $\widetilde\Pi\ne 0$, which in Fig.~\ref{fig:GdeE24} corresponds to the first excited state (eigenenergy around 1 GeV in the figure), is moved beyond the level corresponding to the first excited $\rho \pi$ state. In this case the results with the analysis with stable $\rho$ would be similar to those with the unstable $\rho$. Furthermore, one should also take into account that actual simulations like the one of \cite{sashatalk}, which do not incorporate three-pion correlators, would not see the decay of the $\rho$ into $\pi \pi$, in which case, the analysis with a stable $\rho$ would be more appropriate to interprete such lattice results.

\section{Summary}

We showed how to tackle the problem of the interaction of two particles 
 quantized  in a
finite box of size $L$ when one of the them has a
finite width. The idea is based on extending previously known techniques
for the stable case but quantizing also inside this finite box the decay
channel of the unstable particle. In this way, the continuous integration
needed  to evaluate the selfenergy is substituted by a discrete sum over
the allowed levels in the box of the decay channel of the unstable
scattering particle. We illustrate the method with the $\pi\rho$
scattering which generates dynamically the $a_1(1260)$ resonance within
the chiral unitary approach. The scattering energy levels inside the box
for periodic boundary conditions, both for the stable and unstable case
are compared. The results show significant differences
between both cases.

Then we explain how to solve the inverse problem of getting physical
observables in the real world (infinite volume) from lattice data which
are computed in a finite box. The idea is based
on the improvement of the 
L\"uscher's approach developed in Ref.~\cite{misha}, properly adapted to the
present problem. We apply two methods to solve the inverse problem: the
first one is a previously proposed way based on fitting the parameters
of a given potential to get the  lattice data energy levels, {\it fit
method}.  The second one, {\it direct method}, is to evaluate directly  the $\pi\rho$ amplitude using
Eq.~(\ref{extracted_1_channel2}). The advantage of the {\it direct
method}, which is closer to the  original L\"uscher approach, is that
there is no need to provide a specific shape of the potential
since there is no potential involved. However, the drawback is that the
errors are larger than in the {\it fit method} and that the observables
in the infinite volume can be  evaluated only for those energies for
which there are lattice data.

 With respect to the particular case studied here of the $\pi \rho$
  scattering and the $a_1$ resonance, it is quite instructive to observe that
  the amplitudes obtained are rather peculiar and they do not resemble much
  the shape of a Breit Wigner. This feature will have to be considered in
  future lattice QCD studies.  We showed that using the stable $\rho$, the
  first physical energies accessible were around 1100 MeV. However, the
  proliferation of energy levels due to the quantization of the decay
  products of the $\rho$, making the study with the unstable $\rho$, made a
  wider range of energies available, although the induced phase shifts had
  larger errors using the direct L\"uscher method. In such a case, the
  $fit~method$ proposed here could provide a more efficient method to induce
  the phase shifts with much smaller errors.  

Furthermore, we have also discussed that 
if the starting generated lattice data takes into account
 the decay of the $\rho$
meson into two pions, then the analysis of the inverse problem must be
performed considering also the instability of the $\rho$ meson. Otherwise the
results could be numerically unreliable.

The considerations done in the present work should spur the future
lattice calculations to consider also the discretization  of the decay
channels when unstable particles are involved in the scattering. In between, calculations along the line of \cite{sashatalk}, which can be studied with the stable $\rho$ analysis, produced at other volumes or within a moving frame, could bring extra information on phase shifts in the physical region which would eventually determine the peculiar shape of $\pi \rho$ amplitude in the $a_1$ resonance region, with a strong diversion from a Breit Wigner shape.

\section*{Acknowledgments}
 We would like to thank Michael D\"oring and Sasa Prelovsek for useful discussions. This work is partly supported by DGICYT contracts  FIS2006-03438,
 the Generalitat Valenciana in the program Prometeo and 
the EU Integrated Infrastructure Initiative Hadron Physics
Project under Grant Agreement n.227431.

\end{document}